\documentclass[12pt,preprint]{aastex}
\usepackage{amsmath,amssymb,graphicx,natbib}

\newsavebox{\astrutbox}
\sbox{\astrutbox}{\rule[-5pt]{0pt}{20pt}}

\newcommand{\vv}{{\mathbf v}}
\newcommand{\uu}{{\mathbf u}}
\newcommand{\BB}{{\mathbf B}}
\newcommand{\JJ}{{\mathbf J}}

\newcommand{\ez}{{\mathbf e_z}}
\newcommand{\ey}{{\mathbf e_y}}
\newcommand{\OO}{{\mathbf \Omega}}

\newcommand{\gapprox}{\lower.4ex\hbox{$\;\buildrel >\over{\scriptstyle\sim}\;$}}
\newcommand{\lapprox}{\lower.4ex\hbox{$\;\buildrel <\over{\scriptstyle\sim}\;$}}

\slugcomment{To be submitted to ApJ.} 
\shorttitle{Symmetries, scaling and convergence in MRI  turbulence} 
\shortauthors{Bodo, Cattaneo, Ferrari, Mignone \& Rossi} 

\begin{document} 
 
\title{Symmetries, scaling laws and convergence in shearing-box simulations of MRI driven turbulence} 
 
 \author{G. Bodo\altaffilmark{1},        
         F. Cattaneo\altaffilmark{2}, 
         A. Ferrari\altaffilmark{3},
         A. Mignone\altaffilmark{3},
         P. Rossi\altaffilmark{1} }
  
 \altaffiltext{1}{INAF, Osservatorio Astronomico di Torino, Strada Osservatorio 20, Pino Torinese, Italy}
 
 \altaffiltext{2}{The Computation Institute, The University of Chicago, 
              5735 S. Ellis ave., Chicago IL 60637, USA}

 \altaffiltext{3}{Dipartimento di Fisica Generale, Univesita di Torino, via Pietro Giuria 1, 10125 Torino, Italy} 
 
\begin{abstract} 
We consider the problem of convergence in homogeneous shearing box simulations of magneto-rotationally driven turbulence. When there is no mean magnetic flux, if the equations are non dimensionalized with respect to the diffusive scale, the only free parameter in the problem is the size of the computational domain. The problem of convergence then relates to the asymptotic form of the solutions as the computational box size becomes large. By using a numerical code with a high order of accuracy we show that the solutions become asymptotically independent of domain size. We also show that cases with weak magnetic flux join smoothly to the zero flux cases as the flux vanishes. These results are consistent with the operation of a subcritical small-scale dynamo driving the turbulence. We conclude that for this type of turbulence the angular momentum transport is a proportional to the diffusive flux and therefore has limited relevance in astrophysical situations.
\end{abstract} 

\section{Introduction}
The magneto-rotational instability (MRI) is commonly invoked to explain the origin of turbulence in electrically 
conducting discs \citep{Balbus91}. Although the instability itself is amenable to an analytic treatment, much of what is 
currently known about the nonlinear development of this instability is based on numerical simulations. In 
particular, the efficiency of MRI driven turbulence at transporting angular momentum, which is the crucial 
quantity that ultimately controls the accretion rate,  is known almost exclusively through numerical 
simulations. Because of the difficulties inherent in simulating an entire disc the majority of the numerical 
work has been based on the local model known as the shearing-box approximation \citep{Hawley95}. 
The advantages of the 
shearing-box approximation (SBA) are numerous, the cylindrical geometry of the full disc is replaced with the Cartesian 
geometry of a rectangle, simple periodic and shear-periodic boundary conditions can be applied, and most 
importantly, whatever numerical resolution is available can be deployed to resolving the ensuing turbulence. 
However, a number of issues arise that have led several authors to question the applicability of the  shearing-box approximation (SBA) to the study disc turbulence. \citet{Regev08} have noted some inconsistencies in the formulations of  the SBA with uniform background states, they have also questioned the assumption of locality that underlies the derivation of the SBA. Related concerns about locality were expressed by \citet{Bodo08}. However, the biggest issue about the SBA remains the so-called problem of convergence.

There are two configurations in which MRI driven turbulence is studied numerically; one in which there is a net magnetic flux threading the layer, the other in which there is none. In the former case, if the uniform component of the magnetic field is vertical, say, there is a linearly unstable mode with a well defined vertical wavenumber of maximum growth rate that sets the scale of the instability, in the latter no such state exists, and the MRI must set in as a subcritical instability. Similar considerations apply if the initial field is azimuthal.  What exactly determines the characteristic scale of the turbulence in the no-net-flux case is, at the moment, an open question. The conceptual appeal of the no-flux case is that it offers the possibility of a universal state of MRI turbulence in which the disc becomes self magnetized through dynamo action, so that the angular momentum transport depends on the disc properties but not on the amount of flux threading the disc. Clearly much effort has been devoted to determining if such a universal state exists, and the value of the associated turbulent transport. Simply stated, the problem of convergence is that the angular momentum transport measured in numerical simulations based on the SBA with homogeneous background state and zero mean-flux depends on numerical resolution, and decreases as the resolution increases.

This effect was first noted by \citet{Fromang07} in a series of numerical experiments with the ZEUS code and subsequently confirmed by other authors using different codes \citep[eg][]{Pessah07, Simon09, Guan09}. There are a number of important points that should be made. Most of the evidence for the convergence problem is based on codes with no explicit viscosity or magnetic diffusivity in which the dissipation arises solely from truncation errors. An increase in resolution is interpreted as a decrease in effective dissipation, so that the convergence problem can equivalently be stated as a decrease in the effective transport with decreasing dissipation. Recently, however, \citet{Fromang10} indicated that the problem does not arise when explicit viscosity and magnetic diffusivity are included. Also, the problem is seen in simulations with homogeneous background state and periodic boundary conditions in the vertical. Simulations by \citet{Kapyla10} suggest that the convergence problem is absent  if different boundary conditions are applied in which the magnetic field is vertical on the upper and lower boundaries.  Also, the problem seems to disappear when stratification is included \citep{Davis10}, or the aspect ratio is such that that computational domain consists of a sufficiently tall box (Stone,  private communication). Finally, even when the effect is clearly observed there is no universal agreement about the rate at which the transport decreases with resolution. A number of questions naturally arise: what is the origin of the convergence problem? Is it a numerical artifact associated with the indiscriminate use of ideal solvers? Is it a real physical phenomenon related to subcritial dynamo instabilities? Is it that the shearing-box approximation in its simplest form is too idealized to describe MRI turbulence in a disc? What is the asymptotic scaling of the angular momentum transport with resolution/dissipation, and what can we learn from it?

In this paper we address some of these issues. We revisit numerically the problem of MRI driven turbulence in shearing-boxes with uniform background state and periodic boundary conditions in the vertical. We formulate the problem in a slightly unconventional way that is designed to emphasize the role of symmetries in the shearing-box approximation, and  to make it easier to distinguish between numerical and physical effects. 

In the ideal limit, the shearing-box equations are scale invariant, in the sense  that doubling the number of grid points in a simulation, or doubling the size of the computational domain are entirely equivalent. This gives rise to two equivalent limiting procedures; letting the grid spacing vanish for finite domain size, and letting the domain  size become infinite for finite grid-spacing. 
Although the two limits are equivalent, the latter makes it easier to interpret the results in terms of some classical  results from dynamo theory. Using this formulation we consider the results of high-resolution numerical experiments and reach the following conclusions. The convergence problem is real in the sense that the transport by MRI driven turbulence does decrease with decreasing dissipation and vanishes for vanishing dissipation. The problem is related  to two aspects of the shearing-box approximation with periodic boundary conditions,  one is the  inherent symmetry of the shearing-box approximation , the other  is the absence of an effective inverse cascade in the dynamo solutions. We find that there exists an asymptotic regime in which the effects of the symmetries are  apparent but it is realized at very high resolution, or as we shall see presently, at very large system size, suggesting that most of the simulations described in the current literature are not in that regime. Finally, we argue that even though the shearing-box approximation in its simplest form is probably unable to capture the physics of MRI driven turbulence in discs, it nevertheless gives us very useful clues of  which effects are likely to be important, and how to proceed to improve our models.  

The paper is organized as follows: in the next section we discuss the zero-flux case; we review the standard formulations and propose a slightly different approach that better separates between physical and numerical quantities. We present numerical evidence that suggest the existence of an asymptotic state in which the transport becomes independent of system size. We then discuss cases with small but finite flux, and show that they match smoothly with the zero-flux cases. Finally, we discuss the implications for modeling astrophysical discs.

\section{The zero net flux case} \label{sec_zeroflux} 
We begin by discussing the case with no mean flux. Conceptually, this is the simplest case since for an ideal incompressible fluid, the SBA is completely scale invariant. As we shall see presently, this symmetry plays an important role in determining the asymptotic form of the solutions. 

\subsection{Formulation}

The first step is to cast the SBA equations in dimensionless form. Often, this is accomplished by selecting units that depend on the sound speed, although this choice is convenient from an astrophysical point of view it is not the most suitable to bring out the natural symmetries of the equations. Instead, we begin by considering an incompressible fluid--i.e. a fluid with infinite sound speed, with finite viscosity and magnetic diffusivity, and we shall return to the compressible case presently. A detailed presentation of the SBA can be found in \citet{Hawley95};\citep[see also][]{LL07}. The equations in {\it dimensional} form can be written as   :

\begin{equation}
\frac{\partial \vv}{\partial t} + \vv \cdot \nabla \vv + 2 \Omega \times \vv = \frac{\BB \cdot \nabla \BB}{4 \pi \rho} - \frac{1}{\rho} \nabla 
\left(  \frac{\BB^2}{8 \pi} + P \right) - \nabla \left( 2 A \Omega x^2 \right) + \nu \nabla^2 \vv,
\end{equation}

\begin{equation}
\frac{\partial \BB}{\partial t} + \vv \cdot \nabla \BB - \BB \cdot \nabla \vv = \eta \nabla^2 \BB,
\end{equation}

\begin{equation}
\nabla \cdot \vv =\nabla \cdot \BB= 0
\end{equation}
where $\BB$, $\vv$, $\rho$ and $P$ denote, respectively,   magnetic  field, velocity, density and pressure. The local angular velocity
 $\OO = \Omega \ez$ and shear rate
 \begin{equation}
A \equiv \frac{R}{2} \frac{\partial \Omega}{\partial R}
\end{equation}
are considered constants. For a Keplerian disk $\Omega \propto R^{-3/2}$ and $A = -(3/4) \Omega$.    The velocity can be decomposed as the sum of the base Keplerian flow and the fluctuations 
\begin{equation}
\vv = 2 A x \ey + \uu,
\end{equation}
likewise, the pressure can be  decomposed as the sum of the average and the fluctuations
\begin{equation}
P = P_0 + p.
\end{equation}
With these decompositions, the dimensional shearing box equations become

\begin{equation}
\frac{D \uu}{D t} + A x \frac{\partial \uu}{\partial y} + A u_x \ey + 2 \Omega \times \uu = \frac{\BB \cdot \nabla \BB}{4 \pi \rho} - \frac{1}{\rho} \nabla 
\left(  \frac{\BB^2}{8 \pi} + p \right)  + \nu \nabla^2 \uu,
\end{equation}

\begin{equation}
\frac{D  \BB}{D t} + A x \frac{\partial \BB}{\partial y} - A B_x \ey -  \BB \cdot \nabla \uu = \eta \nabla^2 \BB,
\end{equation}

\begin{equation}
\nabla \cdot \uu = \nabla \cdot \BB = 0,
\end{equation}
with $D/Dt = \partial/\partial t + \uu \cdot \nabla$.   The above system of equations contains only three dimensional parameters: the rotation frequency $\Omega$, the viscosity $\nu$ and the magnetic diffusivity $\eta$.


By adopting the following units of time, length, velocity and magnetic field intensity
\begin{equation}
\tau = \frac{1}{\Omega}  ; \qquad  {\mathcal L} =l_D \equiv  \sqrt{\frac{\nu}{\Omega}} ; \qquad u^* = \sqrt{\nu \Omega} ; \qquad B^* =  \sqrt{\rho \nu \Omega}.
\label{eq:scalings}
\end{equation}

the equations can be written in dimensionless form as 

\begin{equation}
\frac{D \hat \uu}{D \hat t} - \frac{3}{4} \hat x \frac{\partial \hat \uu}{\partial \hat y} -\frac{3}{4} \hat u_x \ey + 2 \ez \times \hat \uu = \frac {1}{4 \pi} \hat \BB \cdot \hat \nabla \hat \BB -  \hat \nabla 
\left(  \frac{\hat \BB^2}{8 \pi} + \hat p \right)  + \hat  \nabla^2 \hat \uu
\label{eq:nondim1}
\end{equation}

\begin{equation}
\frac{D \hat \BB}{D \hat t} -\frac{3}{4} \hat x \frac{\partial \hat \BB}{\partial \hat y}  + \frac{3}{4} \hat B_x \ey - \hat \BB \cdot \nabla \hat \uu + \frac{1}{P_m} \hat \nabla^2 \hat \BB = 0
\label{eq:nondim2}
\end{equation}
where the hatted quantities are dimensionless and 
$P_m=\nu / \eta$ is the magnetic Prandtl number.  It is important to note that the induction equation (\ref{eq:nondim2}) only depends on $P_m$, while the momentum equation (\ref{eq:nondim1}) has no adjustable parameters at all. This is because in the SBA with no mean flux, the diffusive scale $l_d$ is the only intrinsic length that can be defined using the dimensional parameters of the problem. 
Clearly, in order to solve equations (\ref{eq:nondim1} -\ref{eq:nondim2}) one also needs to define a computational domain and suitable boundary conditions thereby introducing an external length-scale  $L$ that characterizes the size of the computational box;  the latter, however,  has no direct physical meaning. In particular, in the case of periodic boundary conditions (shearing periodic in the radial direction), formally one is  considering an infinite domain, with $L$ representing a long wavelength cutoff.  The domain size can itself be nondimensionalized  to define the additional parameter $R \equiv L/{\mathcal L}$.  For fixed magnetic Prandtl number, $R$-- the domain size in units of the diffusive scale --is the only adjustable parameter. 

We note that the non-dimensional form of the equations above is different from what is typically found in the literature. In general, the box size is used as unit of length and the square of the  parameter  $R$ is then identified with the Reynolds number. However, in a turbulent flow, the Reynolds number should be more properly defined in terms of the integral scale of the turbulence. The question then becomes  whether asymptotically, for $R \rightarrow \infty$, in MRI turbulence the integral scale is linked to the box size, or to the diffusive scale, or to a combination of both. We note  that if the solutions become  asymptotically independent of the box size, the system of equations (\ref{eq:nondim1} - \ref{eq:nondim2}) do not depend on any non-dimensional parameters, apart from the Prandtl number, i.e. for fixed Prandtl number there is  a universal solution.  

In the zero net flux case, dynamo processes must effectively maintain  the magnetic field against dissipation, however the magnetic field is itself responsible for generating, through the MRI, the turbulence that sustains it. It is well known  that, in a (small-scale) turbulent dynamo, the magnetic energy is concentrated near the resistive scale. In this case, one expects that the MRI driven velocity  may also be characterized by the diffusive scale. In order for the velocity to have a significant component comparable to the box size an efficient mechanism is required that can transfer energy backwards, from the resistive scale to the largest available scale, i.e. an inverse cascade. Whether such an inverse cascade exists is at the heart of the problem. 

As an aside, we note that the condition for the validity of the SBA is that the characteristic scales of the solution  be much smaller than the box size, lest the use of periodic boundary conditions is not justified. Therefore, if there were cases in which the shearing box results would show  a dependence of the solutions on the box size $R$ as $R \rightarrow \infty$, one should actually reformulate these cases by introducing some global scale characteristic of the whole disc.

One of the main objectives of  MRI studies is to determine the efficiency of the transport of angular momentum by Maxwell and Reynolds stresses.  It is customary to  define the quantity
\begin{equation}
\Sigma \equiv \overline{<u_x u_y - \frac{B_x B_y}  {4 \pi \rho}>}
\end{equation}
that represents the box and time averaged value of the total stresses (hereinafter  overbars denote time averages, while angle brackets denote a box average). This quantity has dimensions of square velocity, thus with our choice of non-dimensionalisation it is measured in units of $\Omega \nu$. For fixed magnetic Prandtl number we can write
\begin{equation}
\Sigma \sim f(R) \Omega^2 l_D^2 = f(R) \Omega \nu.
\end{equation}
Here, $f(R)$ is a dimensionless function that  accounts for the possible dependence of the solutions on the externally imposed box size.  When $R = O(1)$ the box size is comparable to the diffusive scale, and, on general grounds,  one expect a strong dependence of $f(R)$ on $R$. It is not clear, {\it a priori}, what dependence one should expect as $ R \rightarrow \infty$. It is important, however, to be clear about the relationship between the asymptotic form of $f(R)$ and the transport efficiency. In turbulence theory, it is customary to refer to ``turbulent transport" to describe a transport process that becomes independent of diffusion for small diffusion, and to ``collisional transport" to describe one such process that remains proportional to the diffusivities when these become small. Because, here, $\Sigma$ is measured in diffusive units ($\Omega \nu$), if  $f(R) \sim const.$ as $R \rightarrow \infty$ the angular momentum transport is entirely collisional. Conversely, in order for the transport to be ``turbulent" in the sense defined above, $f(R) \sim R^2$ as $R \rightarrow \infty$. In fact, any asymptotic dependence of $f(R)$ weaker than $R^2$, will ultimately lead to a transport that vanishes for vanishing viscosity.

Typically, in the literature, the transport efficiency is measured in terms of the parameter $\alpha$. In the incompressible case this is defined as the total stresses measured in units of $L^2 \Omega^2$ \citep{LL07}; in other words
\begin{equation}
\alpha \equiv \frac{\Sigma}{\Omega^2 L^2} =  f(R)  \frac{1}{R^2},
\end{equation}
since $R^2 = L^2 \Omega / \nu$.
 Again, we recover  the result that in order for $\alpha$ to have asymptotically  finite (constant) value the function $f(R)$ must diverge quadratically as $R$ becomes large. 

When  compressibility is taken into account, an additional physical parameter is introduced, namely the sound speed $c_s$, that can be used to define a new length scale  $H = \Omega c_s$. When vertical gravity is accounted for, $H$, represents the scale height of the  disc.  In studies such as the present one, in which vertical gravity is neglected, this length has no direct  meaning, nevertheless it introduces a new non-dimensional parameter $R_{H} = H/{\mathcal L}$, that must be taken into account in the scaling arguments discussed above. 
As before we can write 
\begin{equation}
\Sigma = F(R,R_H) \Omega \nu,
\end{equation}
where $F(R,R_H)$ is a dimensionless function. For $R_H = O(1)$, again we expect a strong dependence of $F$ on $R_H$. However 
for $R_H$ large, $F$ must become asymptotically independent of $R_H$, since this corresponds to the incompressible limit in which $F$ must approach $f$.

Physically, this corresponds to situations in which $H$ is much larger than the characteristic scales of the solutions. Again, if there is no dependence of these scales on the externally imposed box size $L$, the only available length is ${\mathcal L}$ and the characteristic scales of the solutions will be proportional to ${\mathcal L}$. For constant sound speed and therefore constant ratio $H/L$, the limit $R \rightarrow \infty$ corresponds to $R_H \rightarrow \infty$, and therefore the stress behavior should be the same as in the incompressible case. In this case $\alpha$ is defined as in \citet{SS73}
\begin{equation}
\alpha \equiv \frac{\Sigma}{\Omega H c_s} = \frac{\Sigma}{ c_s^2} 
\label{eq:alpha}
\end{equation}
and we therefore expect again the scaling 
\begin{equation}
\alpha  \sim  \frac{1}{R^2}.
\end{equation}

\subsection{Numerical Considerations}

In most of the MRI studies, as well as in this paper, the analysis of MRI turbulence is performed using ideal codes, where dissipative effects only arise due to  numerical truncation. In general, in this case, it is not known how to write the dissipation term explicitly, however it can be said  that the amount of dissipation depends on the ratio between the local scale of variation of the physical quantities and the cell size. Structures comparable to the cell size are subject to strong dissipation  that  decreases when the scale of the structures increases. Different numerical schemes, however, will behave in  different ways, in particular the reduction of dissipation when  the local scale is  decreased will be sharper for high order schemes and  gentler for lower order ones. In  (\ref{eq:scalings})  we defined the dissipation scale in terms of the viscosity and the shear as $l_D = {\mathcal L} = \sqrt{\nu/\Omega}$.  In the numerical approach one is not able to get an explicit expression for the dissipation scale,  nevertheless  it is known that $l_D$ is related to the cell size $\delta$, in a way  that, however,  depends on the scheme.  In addition,  it is know that $l_D$ (in units of $\delta$)  will be larger for lower order schemes and smaller for the high order ones. Thus, in the non-dimensionalization of  the equations, in principle one should use $l_D$, but in practice one can only use $\delta$.  With the  two dimensional parameters $\Omega$ and $\delta$, it is possible to define the units of time $\tau$, length ${\mathcal L}$, velocity $u^*$ and magnetic field $B^*$  as
\begin{equation}
\tau = \frac{1}{\Omega}  ; \qquad  {\mathcal L} = \delta ; \qquad u^* = \Omega \delta ; \qquad B^* =  \sqrt{\rho} \Omega \delta
\label{eq:scalings2}
\end{equation}

By the same considerations discussed above, the scaling of stresses can be written as
\begin{equation}
\Sigma \sim f(N) \Omega^2 \delta^2
\end{equation}
where $N = L/\delta$ represents the number of grid points and  the $f(N)$ factor accounts for a possible dependence of the solutions on the externally imposed box size.   The scaling for $\alpha$ is correspondingly given by
\begin{equation}
\alpha \sim f(N) \frac{1}{N^2}.
\label{eq:alpha2}
\end{equation}
 If there is no  dependence on the box size, i.e. $f(N)$ is constant, we have $\alpha N^2 = \hbox{const}$; this constant however will be different from scheme to scheme,  since different numerical schemes, as discussed above, will have different relationships between the dissipation length $l_D$ and the cell size $\delta$. The proper unit for $\Sigma$ would be $\Omega^2 l_D^2$, therefore we expect
 \begin{equation}
\alpha N^2 \sim \left(\frac{l_D}{\delta}\right)^2.
\label{eq:alpha3}
\end{equation}
Since $l_D$ is expected to be smaller for high order codes the same should be true for the value of $\alpha N^2$. It should be noted that this discussion ignores possible difference in the effective Prandtl number for the different schemes, that can account for additional residual differences between codes.

   For the compressible case, we can repeat the same reasoning as above: we have an additional parameter $N_H = H/\delta$ that represents the number of grid points over H , 
however the dependence on this parameter should disappear, for constant sound speed, in 
the limit $N   \rightarrow \infty$. 

\subsection{Numerical results}

We have performed a series of compressible isothermal shearing box  simulations for the zero net flux case, with different resolutions and  with three different numerical schemes available in the PLUTO code \citep{PLUTO}. The main difference between the three schemes  is the order of accuracy of the reconstruction, more precisely we employed a scheme with TVD linear reconstruction (second order accurate for smooth flows), a parabolic reconstruction (PPM) and a monotonicity preserving fifth order reconstruction (MP5) (for a description of this scheme see \citet{Suresh97, Mignone10a, Mignone10b}).   Our goal is to  study  the asymptotic behavior of the solutions as the  separation of scales between the box size and the dissipation length becomes large. This can be achieved by increasing the resolution for a fixed scheme, or for a  fixed resolution, by using a scheme with higher order of accuracy.  

In the shearing box approach, the Cartesian coordinates $x$, $y$ and $z$ refer respectively to the radial, azimuthal and vertical directions. Our computational box has aspect ratio $L_x: L_y: L_z = 1 : \pi : 1$.  As in \citet{Fromang07} we use cells elongated in the azimuthal direction, with aspect ratio $1 :  2 : 1$. For each of the schemes we used four different resolutions: 32, 64, 128 and 256 points in the vertical direction, the corresponding grid sizes are therefore $32 \times 50 \times 32$, $64 \times 100 \times 64$, $128 \times 200 \times 128$ and $256 \times 400 \times 256$. The sound speed is chosen such that $H = L$,  the initial magnetic field is $(0, 0, B_0 \sin(2 \pi x))$, with $B_0$ corresponding to $\beta = 1500$ and  random noise in the $y$ component of the velocity is introduced initially to start the growth of the instability. 

  As it is well known, the time histories of  quantities like the box averaged magnetic energy or the box averaged Maxwell or Reynolds stresses, have an initial transient followed by the establishment of a  stationary state in which the quantities fluctuate around a well defined mean value. The amplitude of the fluctuations decreases with increasing resolution. To estimate the mean value we average over the simulation time excluding the initial transient phase. The simulation time varies from case to case and it is always larger than 100 revolutions. We estimate the error of the mean value by subdividing  the averaging interval in ten subintervals and then computing the standard deviation of the subinterval averages. The resulting estimate of the error is always less than 10\%.

\subsubsection{Transport}

\begin{figure}[htbp]
   \centering
   \includegraphics{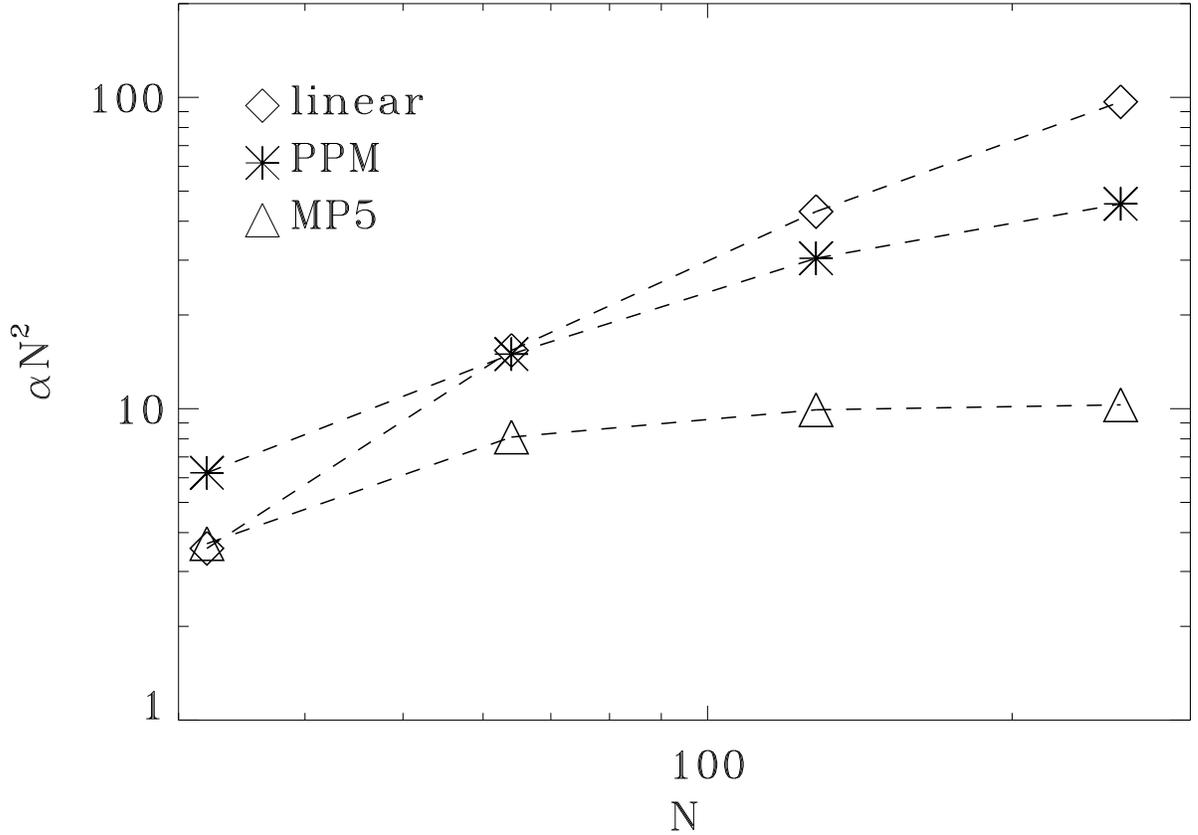} 
   \caption{Plot of $\alpha N^2$ as a function of $N$.  The three curves refer to the three different numerical schemes used in the simulations. These  are  identified by the symbols in the legend. In all cases, the error bars are smaller than the symbols. The slope is about 1, corresponding to the results by \citet{Fromang07} with the ZEUS code, for the PPM curve between N=64 and N=128 .}
   \label{fig:maxw}
\end{figure}

We start our analysis by considering the behavior of $\alpha$, as defined in  (\ref{eq:alpha}), as a function of resolution. Figure \ref{fig:maxw} shows $\alpha N^2$ as a function of $N$ for all the cases  considered. The  quantity displayed represents the function $f(N)$ defined in (\ref{eq:alpha2}), which accounts for a possible dependence of the solutions on the externally imposed box size.  Consonant with our objectives,  we are interested in the asymptotic behavior of $f(N)$ as $N \rightarrow \infty$.  

We recall that in order for  $\alpha$ to be resolution independent, $f(N)$ should diverge as $N^2$. The results by \citet{Fromang07}  suggested that $f(N) \propto N$. A constant value of   $f(N)$, on the other hand,  means that the solution is  asymptotically independent of the box size and that the scaling of $\alpha$ is determined solely by the equations.

In Figure \ref{fig:maxw}, the  squares refer to the linear scheme, the stars to the PPM and the triangles to the MP5.  It is obvious that, as the  the order of accuracy of the scheme is increased, the curves tend to be flatter for increasing $N$.  However, while the linear and PPM schemes show a continually decreasing slope and do not seem to have reached yet an asymptotic behavior, the MP5 scheme, by contrast, appears  to be leveling out.   The results of MP5  suggest that $f(N)$ approaches a constant value as $N \rightarrow \infty$ and therefore that $\alpha$ has the asymptotic natural scaling $\sim 1/N^2$ arising from the equations  alone. Our expectation is that, given sufficient resolution, a similar behaviour would be observed with the other schemes. 

It is important to note that even if all the curves corresponding to different schemes eventually level off, they will not in general have identical asymptotic values. The reason for this is that we do not know the explicit form of the numerical dissipation for each scheme; accordingly we have used the cell size instead of the dissipation length to non-dimensionalize the equations. Indeed, we  note that in agreement with (\ref{eq:alpha3}) the higher order scheme MP5 has a lower  value of the non-dimensional stresses than the other schemes. 

\subsubsection{Structures}

\begin{figure}[htbp]
   \centering
   \includegraphics{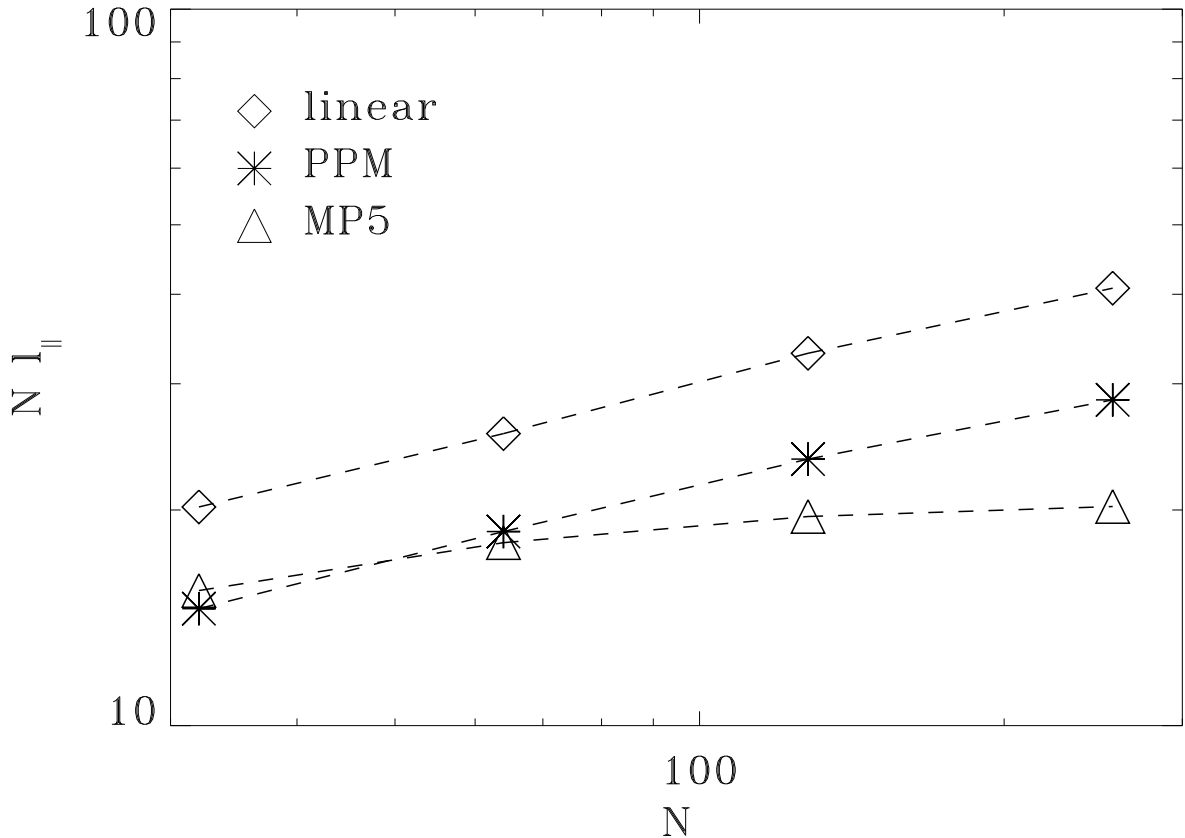} 
   \caption{Plot of $N l_\parallel$ as a function of $N$.  The three curves refer to the three different numerical schemes used in the simulations. These are identified by the symbols in the legend.}
   \label{fig:kpar}
\end{figure}

\begin{figure}[htbp]
   \centering
   \includegraphics{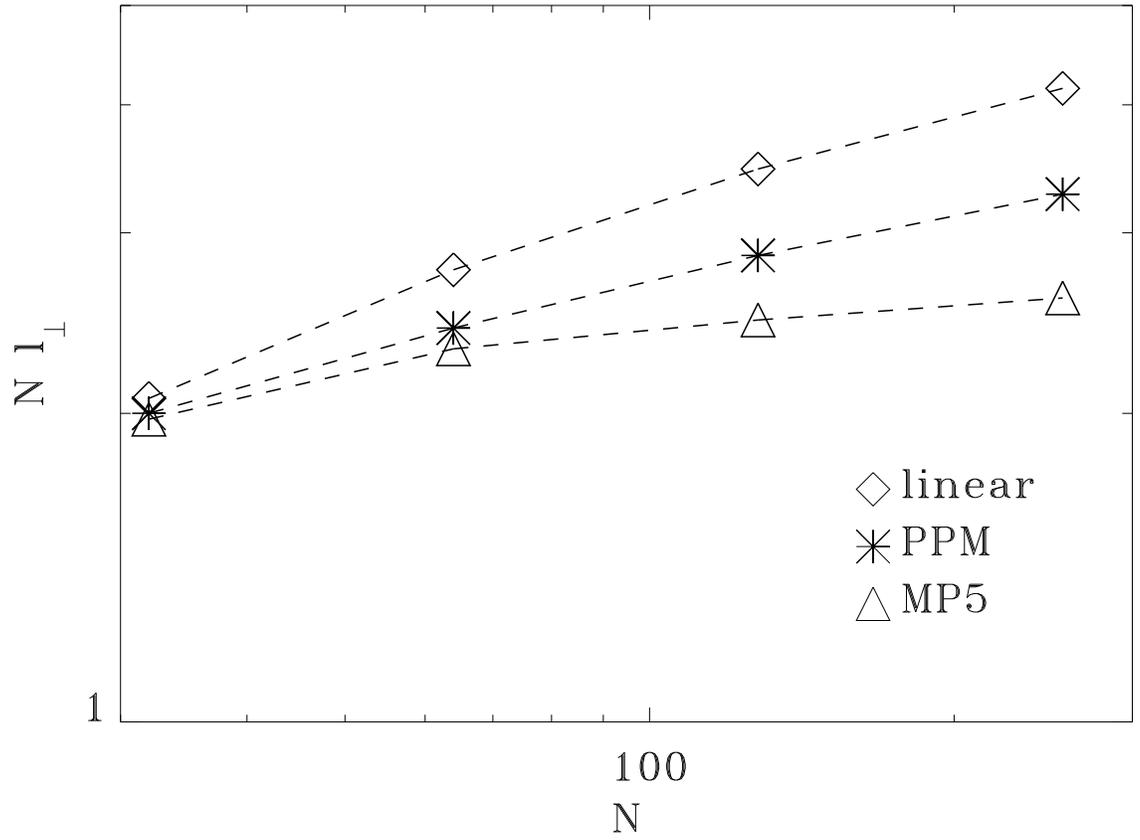} 
   \caption{Plot of $N l_\perp$ as a function of $N$.  The three curves refer to the three different numerical schemes used in the simulations. These are identified by the symbols  in the legend.}
   \label{fig:kperp}
\end{figure}

The behavior of $\alpha$ as a function of resolution suggests that the  only scale that determines the properties of the solutions is the dissipative scale, which, in numerical studies, is related to the cell size. It is therefore important to examine the behavior of the  size of the typical observed structures as a function of resolution. In order to characterize the scales of magnetic structures, we measure the average  scales of variation of the field  in the directions parallel and perpendicular to itself. We  define  the quantities 
\begin{equation}
l_\parallel = \overline{\left( \frac{< | \BB |^4>}{<|\BB \cdot \nabla \BB|^2>}  \right)^{1/2}}; 
 \qquad l_\perp = \overline{ \left( \frac{< | \BB |^4>}{< | \JJ \times \BB |^2>} \right)^{1/2} }
\end{equation}
which represent measures of the characteristic scales of the magnetic field respectively in the parallel direction and along the direction of maximum gradient \citep[see][]{Schekochihin04}.
In fig. \ref{fig:kpar} and \ref{fig:kperp} we plot respectively $N l_\perp$ and $ N l_\parallel$ as  functions of $N$. As before, the three curves refer to the three different numerical schemes and the symbols have the same meaning as in  the previous figure. The lengths are normalized with respect to the cell size and the plotted quantities therefore represent the average number of grid points on which the observed structures extend. As expected, the magnetic structures are highly anisotropic with $l_\parallel / l_\perp \sim 10$. Considering the dependence on resolution,  we see again  that the linear and PPM schemes show an increase of the number of grid points over which magnetic structures extend, while the MP5 scheme appears to tend to a constant number of grid points independent of resolution, both in the longitudinal and transverse directions.        

This analysis suggests that the magnetic field is characterized by  elongated  structures extending  transversally over a few grid points. In order to get further insight into the three-dimensional structure one can define a third scale in the direction orthogonal to both $\BB$ and $\BB \times \JJ$  and that points approximately in the vertical direction:
\begin{equation}
l_{\BB \cdot \JJ} = \overline{ \left(  \frac{< | \BB |^4>}{<|\BB \cdot \JJ|^2>}  \right)^{1/2} }.
\end{equation}

\begin{figure}[htbp]
   \centering
 \includegraphics{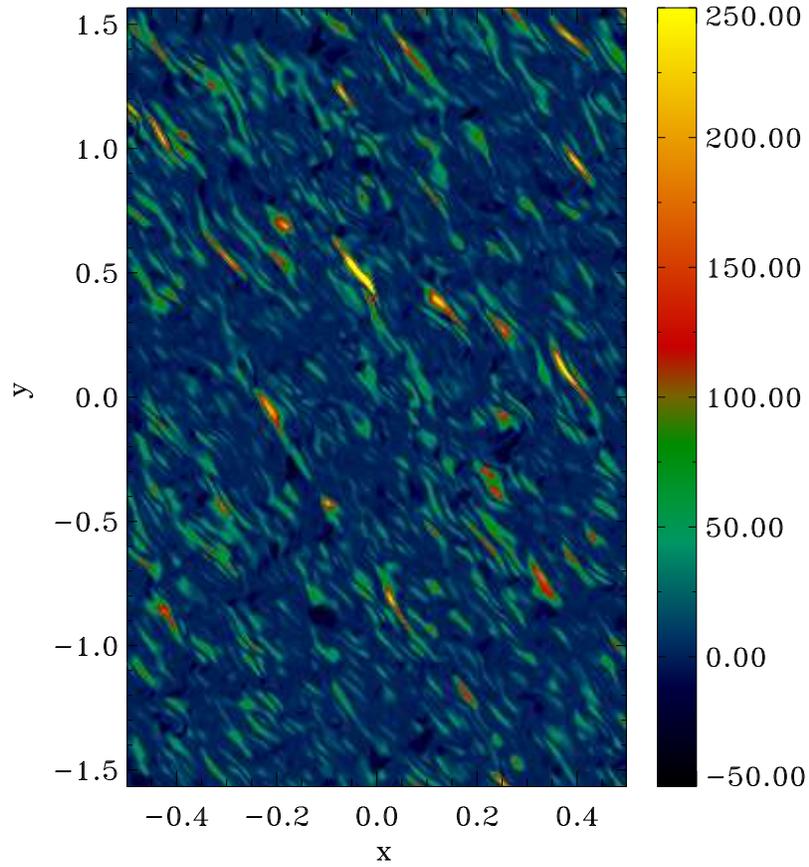} 
   \caption{2D cut  in the $x-y$ plane of the distribution of Maxwell stresses. The distribution is characterized by thin filaments with high stress values separated by wide regions where the stresses are almost zero.  }
   \label{fig:imgstr}
\end{figure}

The ratios of the scales along the three directions are tipically $l_\perp : l_\parallel : l_{\BB \cdot \JJ}  \sim 1 : 10 :  5$ indicating that the magnetic structures can be thought of as thin sheets. This result is also discussed by \citet{Guan09}, who base their analysis on the correlation function. They find that  the correlation lengths scale as $N^{-2/3}$ and that therefore, measured in units of the cell size, show an increase with resolution, suggesting that one could expect some kind of transition. We have also repeated this convergence analysis on the correlation lengths and again we find that while the linear and PPM schemes show a behavior similar to that discussed by Guan et al. (2009), the MP5 scheme at a resolution of 256 grid points shows a convergent behavior, i.e. the correlation lenghts scale as the cell size.  A better impression of the magnetic field structure can be obtained by looking at  Fig. \ref{fig:imgstr}, where we show a 2D cut of the Maxwell stress distribution in the $x-y$ plane at $z=0$ . The 3D sheets  appear in the figure as filaments of high Maxwell stresses  and high magnetic field intensity, separated by regions of low magnetic field.

 We can now discuss this filamentary  structure from a more physical point of view.     
 From the y-component of the induction equation, it can be seen  that the field structure is determined by the competition between three terms: the stretching by the background shear, the stretching due to velocity fluctuations ($\BB \cdot \nabla \vv$ term) and dissipation, and the transverse size is determined by the balance between these terms. The background shear tends to stretch the field along the azimuthal direction, simultaneously decreasing its perpendicular scale of variation, until dissipation sets in. These magnetic filaments, however, give rise to a local transport of momentum and, therefore, tend to reduce the local shear. One, therefore expects that the effect of the nonlinear term $\BB \cdot \nabla \vv$ is to reduce the stretching by the background shear. The observed increase of  the normalized transverse scale with increasing resolution may be related to the increase of this effect, which however, ultimately, has to be limited since it cannot completely suppress the background shear. Thus, asymptotically the transverse scale has to remain constant and be  determined by the balance between shear stretching and dissipation.  The MP5 scheme, which is  closer to the asymptotic behavior, accordingly shows a tendency of the transverse scale to become constant. This leads to the conclusion that a possible dependence on the scale of the box may appear  only in the parallel scale.  However, the ratio $l_\parallel / l_\perp$ seems to be independent from resolution,  i.e. the magnetic structures  do not appear to become more elongated as  the resolution increases and  all the characteristic lengths appear to become asymptotically  constant when measured in unit of the dissipative length.     

\subsubsection{Probability Density Functions}
       
\begin{figure}[htbp]
   \centering
 \includegraphics{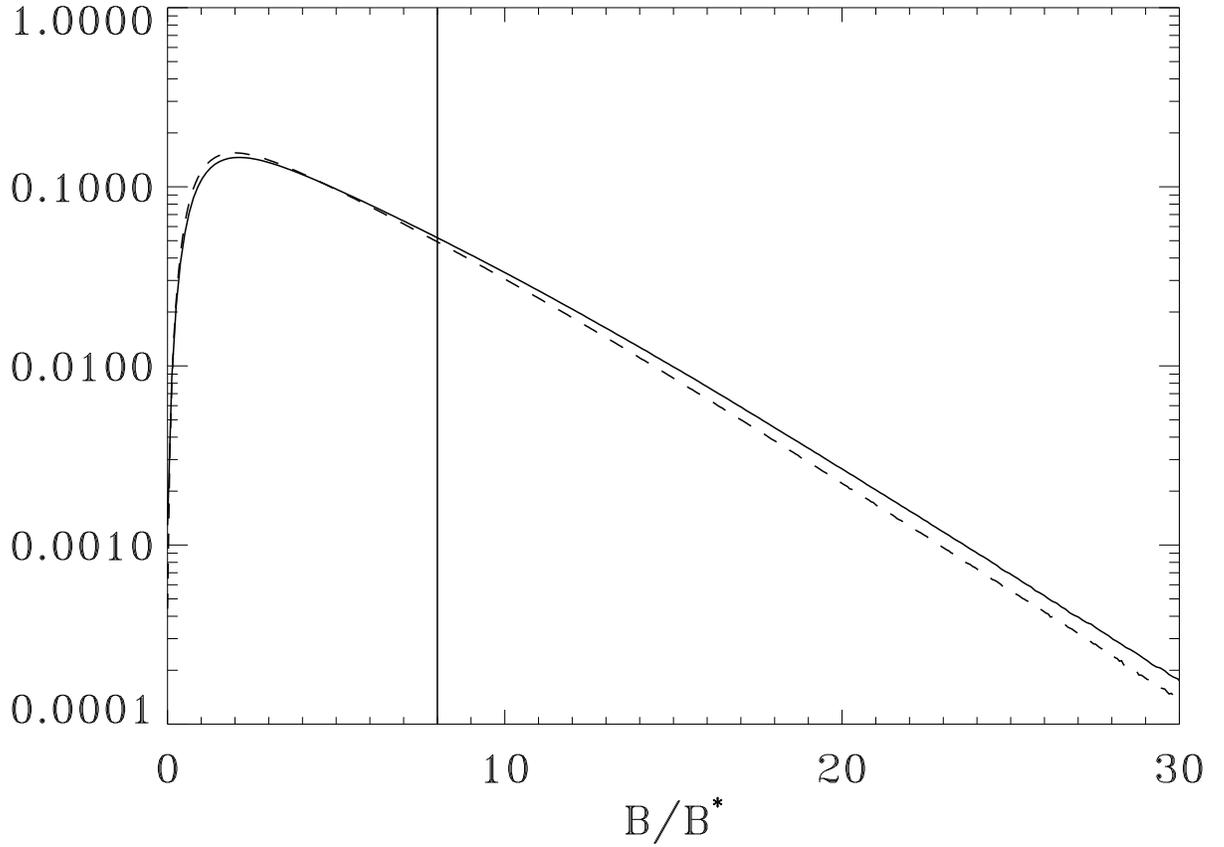} 
   \caption{Probability distribution functions of the modulus of the magnetic field measured in units of $B^*$ for two MP5 cases with different resolutions. The number of grid-points per unit length--measured in units of $L$--are  128 (solid line) and 256 (dashed line). The field values larger than that indicated by the vertical line contribute to about $70\%$ of magnetic energy and $80\%$ of Maxwell stresses.}
   \label{fig:pdf}
\end{figure}

 Further details on the solutions can be examined  by considering  the behavior of the probability distribution function (pdf) of magnetic field intensity and  of the second order structure functions. Our purpose is twofold: on one hand we want to show that, indeed, if we measure all quantities in their proper units, there is a convergence to a well defined solution when we increase the number of grid points and, on the other hand, we want to better characterize this solution. We start by examining the probability distribution function (pdf) of magnetic field intensity.  Figure \ref{fig:pdf}  compares the two MP5 cases with the highest resolution, the pdf's are normalized to unity and the field values are measured in units of $B^*$. 

The pdf's  have an exponential character, the mean value of $| \BB | /B^*$ is about 5.5 with a difference of less than $1\%$ between the two cases at different resolution. The contribution to the magnetic energy and to the Maxwell stresses comes  mainly from the high field values;  values to the right of the vertical line plotted in the figure contribute to about $70\%$ of the total magnetic energy and to about $80\%$ of total Maxwell stresses, while occupying only about $20\%$ of the volume. The higher contribution to the Maxwell stresses with respect to magnetic energy is explained by the fact that these higher field values correspond to the magnetic structures discussed above. For these,  there is a strong correlation between the radial and azimuthal components of the field, while for lower field values the correlation is  much lower. The fact that the volume fraction supporting the transport is independent of resolution explains the decrease in the amplitude of the fluctuations when resolution is increased. In fact, as resolution is increased and the volume fraction remains the same,  the number of points that  contributes to the average gets larger and therefore the fluctuations of the average become smaller.

\begin{figure}[htbp]
   \centering
\includegraphics[width=14cm]{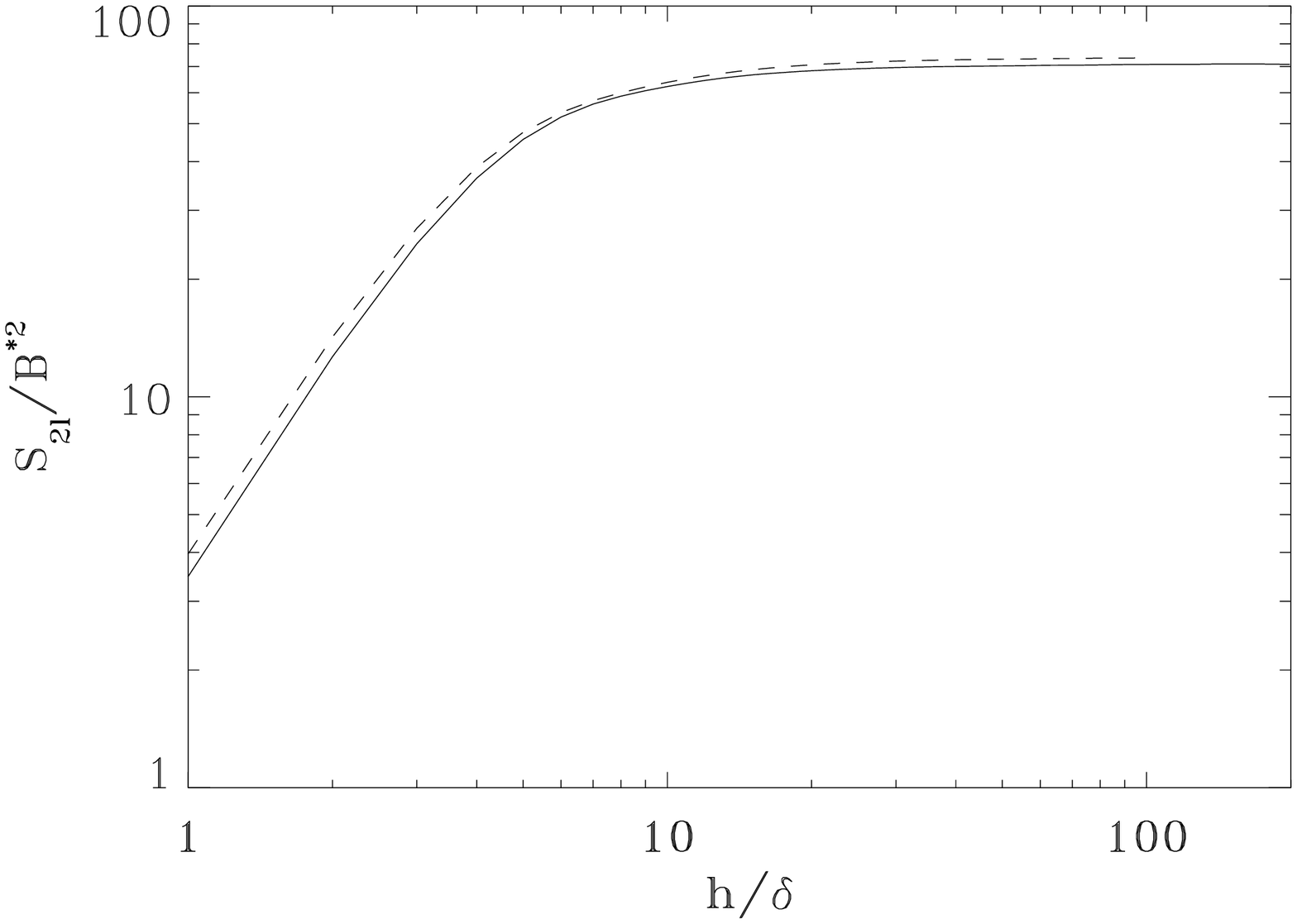} 
 \includegraphics[width=14cm]{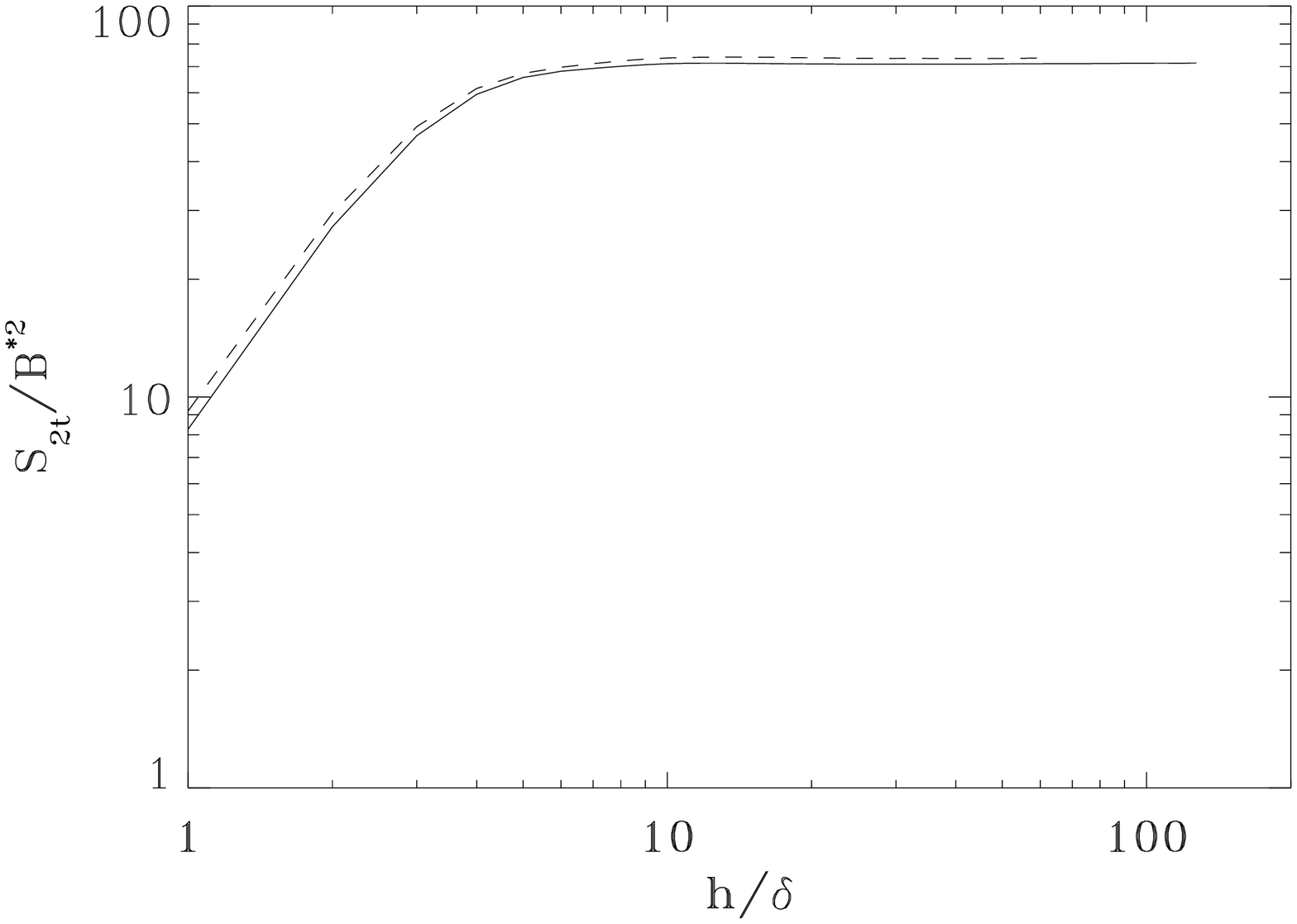} 
   \caption{Longitudinal (left panel) and transverse (right panel), time averaged, second order structure functions of $B_y$. The two curves correspond to two MP5 cases with different resolutions. The number of grid-points per unit length--measured in units of $L$--are  128 (solid line) and 256 (dashed line)}
   \label{fig:struct1}
\end{figure}

In fig \ref{fig:struct1} we plot the time averaged second order longitudinal and transverse structure functions for the $y$ component of magnetic field, for the two MP5 cases at the highest resolution. They are explicitly defined as
\begin{equation}
\overline {S_{2l}} = \overline{< \left( B_y (x, y+h, z) - B_y(x,y,z) \right)^2 > }     \qquad
 \overline {S_{2t}} = \overline{< \left( B_y (x+h, y, z) - B_y(x,y,z) \right)^2 > } 
\end{equation}
 Following our discussion, we compute $S_{2l}$ and $S_{2t}$ in units of $B^{*2}$ and $h$ in units of $\delta$. Again, it can be seen  that, using these units, the difference between the two cases at different resolution is very small, indicating a convergent behavior. The structure functions in both directions become flat at scales larger than few grid points. This plateau indicates that the values of magnetic fields become uncorrelated at these scales. A similar conclusion could be reached from the behavior of the averaged spectra, like those shown in \citet{Fromang07}, that become flat at small wavenumbers. 

From these two results, we  conclude that the solution is characterized by the superposition of uncorrelated magnetic sheets. Those with high field intensity also have a high degree of correlation between magnetic field components, and consequently give the highest contribution to both the total magnetic energy and Maxwell stresses. An increase of the domain size does not change the properties of these sheets, but  simply increases the available volume and therefore the number of sheets, leaving unchanged all the average properties. In a similar way there should not be any modification of the solution by changing the aspect ratio of the computational box, and we have indeed tested this by examining two further cases with aspect ratios respectively $1:2\pi:1$ and $2:\pi:1$, with MP5 and $128$ points in the vertical direction, finding no changes in all the averaged properties.

\section{The finite net flux case} \label{sec_netflux} 
We now turn to the case in which a net magnetic flux threads the computational domain.
In this case the magnetic field has a uniform component of strength ${B_0}$, say, that can be used to define a new dimensional length scale  
 $\lambda_A$ given by
\begin{equation}
\lambda_A = \frac{1}{\Omega} \frac{B_0}{\sqrt{4 \pi \rho}}.
\end{equation}
Accordingly, the averaged stresses will show a dependence also on this scale and keeping the same units as in the previous section, in general we can  write
\begin{equation}
\Sigma = g \left( \frac{\lambda_A}{l_D},  \frac{L}{l_D} \right) \Omega^2 l_D^2.
\end{equation}
Here,  we are mostly interested in the form of the function $g$, when $L/l_D = R \rightarrow \infty$. On general grounds,  we  expect, it will take different forms for different relationships between the three scales $L$, $\lambda_A$ and $l_D$. We start our discussion by considering what is known from numerical simulations in the available literature.  The case that has received the greatest attention is that of a mean vertical field; the results have been summarized in  \citet{Pessah07}. For this case the system is linearly unstable to the MRI; the (vertical) wavelength of maximum growth rate is given by $\lambda_M = 2 \pi \sqrt{16/15} \lambda_A$, the instability exists only for wavelengths $\lambda > 1/\sqrt{3} \lambda_A$. If  $\lambda_A$ is larger  than  $\sqrt{3} L$, no unstable wavelengths fit in the box, the system becomes stable, and the stresses drop to zero. \citet{Pessah07} show that as $\lambda_A$ increases, the stresses scale linearly with $\lambda_A$,  reach a maximum and, eventually, drop to zero when the above condition is met. 
Furthermore,   it has been found   that, at fixed $\lambda_A/L$ that the stresses increase with resolution ---i.e. with $R$ \citep[see e.g.][]{Silvers08, Bodo08}; thus, for high enough resolution, they  should  converge to a value independent  of $R$.  Also, we have  to note that \citet{Bodo08} have found a dependence of the solutions on the aspect ratio $L_x/L_z$ of the computational box, with a convergent behavior for high enough values of $L_x/L_z$. Putting these considerations together, we conclude that for  $l_D \ll \lambda_A \lesssim L$,

\begin{equation}
\Sigma \sim g \left( \frac{\lambda_A}{L} \right) \Omega^2 L^2,
\label{eq:meanfield1}
\end{equation}
where the function $g$ is initially proportional to $\lambda_A/L$, then reaches a maximum (at a fixed value of $\lambda_A / L$) and finally  drops to zero for $\lambda_A > \sqrt{3} L$. Thus the  value of $\Sigma$ at the maximum  scales as $\Omega^2 L^2$.

In the opposite regime, when $l_D \sim \lambda_A << L$,  the averaged stresses tend towards the value obtained in the zero flux case \citep{Pessah07}.  The results of the previous section  suggest that, for large $R$,  $\Sigma$ is independent of $L$ and scales as $\Omega^2 l_D^2$. 

\begin{figure}[htbp]
   \centering
   \includegraphics{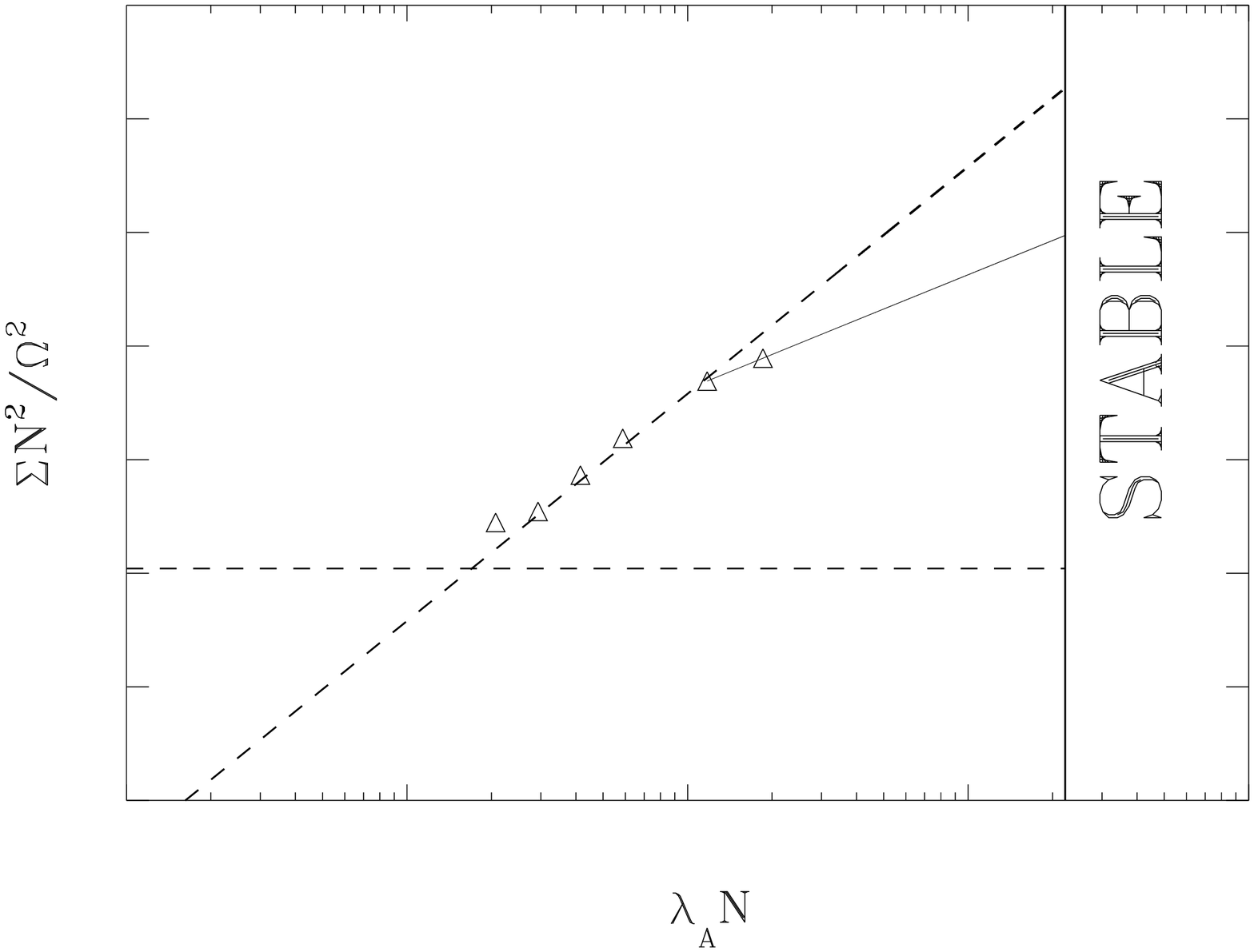} 
   \caption{Plot of the stresses as a function of $\lambda_A$, the triangles represents the results of numerical calculations carried out with the MP5 scheme, with 128 points in the vertical direction. The horizontal dashed line represents the averaged stress value for the zero flux case; the dashed line shows a quadratic behavior; the solid line shows a linear behavior and the vertical solid line shows the linear stability boundary.}
   \label{fig:scalingnetflux}
\end{figure}

In order to connect smoothly the two portions with, respectively, $l_D \sim \lambda_A \ll L$ and  $l_D <<  \lambda_A \lesssim L$, there must be an intermediate range of values of $\lambda_A$ in which the stresses scale as $\Omega^2 \lambda_A^2$. In this range the solution becomes insensitive both to $l_D$ and to $L$, this can be achieved only if  $\lambda_A$ is sufficiently far from both scales and therefore the extent of this intermediate range with quadratic scaling should increase as $R$ increases. Contrariwise, decreasing $R$  should cause the interval to shrink eventually to disappear for low values of $R$.  In fig. \ref{fig:scalingnetflux},  we show the results of calculations with a net vertical  flux performed at a resolution of 128 points in the vertical direction, with the MP5 scheme. The horizontal dashed line corresponds to the averaged stress value for the zero flux case, the dashed line corresponds to a quadratic scaling, the solid line corresponds to a    linear scaling and the vertical solid line marks the stability boundary; to the right of it, the stresses vanish.

It is apparent from the figure, that  indeed, for small enough $\lambda_A$,  a range  of values of  exists with quadratic scaling, while for larger $\lambda_A$ the curve reverts the to linear scaling  discussed above.  

We have deliberately restricted our investigation to   values of $\lambda_A$ for which we have quadratic scaling and the transition from quadratic to linear. 

Larger values of $\lambda_A$, for which the stresses increase linearly up to a maximum and drop to zero to the right of the stability boundary have already been extensively discussed in the literature  \citep[see e.g.][]{Pessah07}.
We note here, that the portion with quadratic scaling has never before been reported in the literature (however \citet{LL10} find evidence of a scaling steeper than linear), this is most likely due to  limited resolution since, as discussed above, for low values of $R$, the quadratic portion  disappears.  
We believe that the reason why we were able to detect it is because of the high order of accuracy of the MP5 scheme, which for the same resolution, produces a much higher {\it effective} value of  $R$. 

\section{Discussion and Conclusion} 
In this paper, we have addressed the problem of convergence in numerical studies of MRI driven turbulence in the homogeneous shearing box approximation. Any simulation that is physically useful should have some meaningful property that becomes independent of the number of grid points as the latter becomes large. This does not seem to be the case here.  As noted by several authors, in the weak flux regime--i.e. when the total magnetic flux threading the computational box is small or zero, the dimensional angular momentum transport 
decreases as as the resolution increases eventually becoming arbitrarily small \citep{Fromang07,Pessah07, Simon09, Guan09}. There are at least two possible frameworks in which to address this issue: one in terms of Reynolds number, the other in terms of computational domain size. In both cases one is interested in the asymptotic behavior of the solutions as either the Reynolds number or the domain size become large. Clearly, both frameworks are related and to some extent it is a matter of preference which one it is picked. In this paper we have chosen to discuss the convergence problem in terms of domain size. The reason is that, because of the symmetries of the homogeneous shearing box approximation, there is no a priori characteristic outer scale for the velocity, which leads to an ambiguity in the definition of the Reynolds number. On the other hand there is a natural inner scale--i.e. the dissipation length--that can be used to non dimensionalize the equations. In this case, the Reynolds number is fixed, and of order unity, and the only remaining free parameter is the domain size. With this choice, increasing the resolution and increasing the domain size are entirely equivalent operations. It is then possible to define the characteristic scale of the solutions, once they are computed, in terms of, say, the (inverse) scale at which the velocity or magnetic spectrum peaks; the latter quantity being conceptually  useful since it is related to the effective angular momentum transport. The problem of convergence then relates to the behavior of the characteristic scale of the solutions as the computational domain size increases. 

As we discussed in section 2, there are two limiting possibilities: one in which the solution scale diverges with the domain size, the other in which it becomes asymptotically independent of it. The former can lead to a net turbulent transport that is independent of diffusivity, the latter to a quasi-collisional transport that scales as some fixed multiple of the  diffusivity.
The evidence from the numerical simulations, in particular figure \ref{fig:maxw}, is that  it is the latter and not the former that gives the correct asymptotic behavior. In other words, the characteristic scale of the solutions becomes independent of the size of the computational domain. If we accept this result as correct, there are a number of issues that naturally arise. 

The first concerns the physical interpretation of the result. In the weak field regime, MRI driven turbulence should be considered within the framework of a subcritical  dynamo instability since the magnetic field necessary for the development of the MRI must be generated self-consistently by the MRI turbulence itself.
In dynamo theory, it is customary to distinguish between two types of dynamo action, small-scale and large-scale. Small-scale dynamos generate magnetic field with characteristic scale comparable to, or smaller than the characteristic scale of the velocity field. The only requirements for the operation of such dynamos are that the magnetic Reynolds number be sufficiently high, and that the underlying velocity not be too symmetric. Clearly, both conditions can be satisfied here provided the system size is large. 
Large-scale dynamos, on the other hand, generate magnetic field with characteristic scale larger than that of the velocity. The latter types of dynamo are  associated with flows lacking reflectional symmetry, i.e. helical flows, and an inverse cascade of magnetic helicity. Recently, \cite{TCB11} have argued that in the nonlinear regime it is often difficult to distinguish unambiguously between large- and small-scale dynamo action, and have introduced instead  the idea of a system-scale dynamo, as one that produces magnetic structures comparable in size to the system scale irrespective of the scale of the velocity. The fact that, here, the scale at which the magnetic spectrum peaks is comparable with the dissipation scale and it is asymptotically independent of the system size indicates that the dynamo operating here is of the small-scale type \citep{VK86, BC04}. Dynamos of this type have no manifest inverse cascade of (magnetic) helicity, they can lead to the production of magnetic energy but not of substantial magnetic flux. In the specific context of MRI driven turbulence a small-scale dynamo will not give rise to a ``turbulent" transport of angular momentum. When regarded in the general scheme of dynamo theory, this result is actually not that surprising since the underlying flow is not strongly helical, and therefore, there is no a priori reason to expect an efficient inverse cascade. 

This brings us to the second issue, which relates to the astrophysical significance of these types of MRI solutions. Clearly, it is to be expected that this type of dynamo excitation will be prevalent in most (electrically conducting) discs. Its role in the transport of angular momentum, and therefore in regulating the accretion rate, however will be negligible. With things as they stand, the only dynamo solutions that might have an impact on the angular momentum transport are those with an efficient inverse cascade. Interestingly, this implies that by their very nature such solutions should not be treated in the context of a local approximation.  We find ourselves in a paradoxical situation in which if a solution is astrophysically relevant it cannot properly be described by a local  model, and if it can be described by a local model then it is astrophysically irrelevant. 
The third issue concerns the role of numerics in influencing what we understand, or think we understand in computational astrophysics. It is interesting that the asymptotic behavior that eventually emerges according to the present study is in many ways the most natural: that consonant with a subcritical small-scale dynamo instability developing in a system where there are not many reasons to expect otherwise. 
Likewise, the scaling behavior for the cases with weak imposed fluxes is asymptotically the one that makes most sense given the symmetries of the equations. Yet, their numerical realizations were somewhat long in the making, and had not been reported in the literature to date. The reason is that these asymptotic regimes requires a substantial dynamic range to manifest themselves. In this particular case, this was achieved by a combination of high resolution and high accuracy numerics. If we had tried to obtain the correct limiting behaviors with the less accurate codes, it would have been prohibitive\footnote[1]{It is useful to note that the scheme with quadratic reconstruction used here has order of accuracy comparable with ZEUS and ATHENA; two of the codes commonly used by the practitioners of numerical studies of the MRI}.
Finally, it behooves us, to compare the conclusions of the present study with those of related studies in which different scaling behaviors have been observed. Of particular relevance   are the work of \citet{Kapyla10} who also consider unstratified shearing boxes but with non periodic boundary conditions; the work of \citet{Fromang10} who has the same set up as ours but with explicit diffusivities; and the work of \citet{Davis10} who include stratification. In all these cases the authors report that the problem of convergence as stated above, is absent. In the present context, these conclusions can be attributed to one or more of the following possibilities: the simulations are not in the asymptotic regime because of limited resolution and/or accuracy; the extra physics has engendered an inverse cascade--i.e. the type of dynamo action has changed from small-scale to  large-scale; the extra physics has changed the nature of the subcritical dynamo instability and has caused it to become of the system-size type. The last two possibilities are related but distinct. They both refer to circumstances that lead to the production of large scale magnetic fields, but in a large-scale dynamo the velocity correlation length remains small, it is only the magnetic structures that become large; in a system-scale dynamo both the velocity and magnetic structures grow in size until they occupy the largest available scale--the system size--hence the name
\citep{TCB11}.   

We begin by discussing the work of \citet{Kapyla10}. They consider unstratified shearing box simulations with boundary conditions at the upper and lower boundaries corresponding to impenetrable, stress-free velocities and vertical magnetic fields. The distinctive feature of these conditions is that they allow a flux of magnetic helicity in and out of the box. This has been advocated by several authors as a feature that might be beneficial to large-scale dynamo action \citep[see e.g.][]{VC01}. The calculations are carried out with explicit diffusivities and the authors report that the effective angular momentum transport is substantial and  insensitive to variations in the values of both the magnetic Reynolds and Prandtl numbers. The morphology of the solutions is dramatically different from the periodic cases, with two large current boundary layers forming in the vicinity of the upper boundaries and the generation of a coherent nearly uniform azimuthal field. All these findings are consistent with the activation of a large-scale type dynamo. There remains to be seen if these solutions persist at higher resolutions, as well as how they map to a global geometry. These issues notwithstanding, the \citet{Kapyla10} solutions are quite remarkable and could potentially have considerable astrophysical significance. 
Next, we discuss stratification \citep{Davis10}. Its presence drives two new effects: it introduces a new spatial scale in the problem, namely the vertical scale height that breaks the symmetry of the homogeneous, periodic models, and it also introduces buoyancy forces associated with thermal, pressure, and magnetic pressure fluctuations. It is not unlikely that these new effects could modify the type of dynamo action and drive flows and field coherent on scales comparable with the scale height. If this is indeed the case, it would be interesting to see how the solutions behave as the simulation size is increased for fixed scale height. Such a study would be numerically very demanding but astrophysically very useful.

Finally, we consider the recent results of \citet{Fromang10}. They report that when explicit diffusivities are included, the angular momentum transport becomes independent of resolution--i.e. the convergence problem goes away. If this result has a physical basis, it implies that the presence or absence of an inverse cascade and the type of associated dynamo action (large-scale as as opposed to small-scale) depends on the detailed form of the diffusivities. The dynamo is small-scale for all three types of numerical diffusivities considered here and in other similar studies, and large-scale for ``physical" diffusivities. Although this possibility is not inconceivable, it is somewhat peculiar. It implies that the type of dynamo action is determined not by the symmetries of the problem, or large-scale features, but by the micro-physics. If correct it is a remarkable result. Another possibility, is that the simulations are not in the asymptotic regime. To explore this possibility we note that the study by \citet{Fromang10} was carried out by incorporating explicit diffusivities into the ZEUS code. In order for the explicit diffusivities to be correctly represented they must give rise to boundary layers whose size is not smaller than those arising from truncation errors. Thus the dynamic range of a code with numerical diffusivities gives an upper bound on the dynamic range achievable with the same code, the same resolution and explicit diffusivities. Now, ZEUS has an order of accuracy comparable to the  code with quadratic reconstruction, thus we can ask the following question. What is the resolution that we need with the quadratic code to see the same asymptotic behavior we observe with the MP5 code? This can be estimated by comparing the curves in Figure \ref{fig:maxw} and taking the ratio in resolutions for which the curves corresponding to the PPM and MP5 code have similar slopes. This exercise gives a factor close to 4. A more careful analysis comparing boundary layers in a channel flow problem yields a similar estimate. Thus our conclusion is that our quadratic code requires a resolution exceeding 1000 grid-points to reproduce the asymptotic behavior observed by the MP5 code with 256 grid-points. Since the highest resolution in the work by \citet{Fromang10} was 512 grid-points it is possible that their results may not yet be in the asymptotic regime.

\section{Acknowledgment}
This work was supported in part  by the National Science Foundation 
sponsored Center for Magnetic Self Organization at the University of Chicago. 
Calculations were performed at CINECA (Bologna, Italy) 
thanks to  support by INAF and at CASPUR (Rome, Italy).





\begin{thebibliography}{24}
\expandafter\ifx\csname natexlab\endcsname\relax\def\natexlab#1{#1}\fi

\bibitem[{{Balbus} \& {Hawley}(1991)}]{Balbus91}
{Balbus}, S.~A., \& {Hawley}, J.~F. 1991, \apj, 376, 214

\bibitem[{{Bodo} {et~al.}(2008){Bodo}, {Mignone}, {Cattaneo}, {Rossi}, \&
  {Ferrari}}]{Bodo08}
{Bodo}, G., {Mignone}, A., {Cattaneo}, F., {Rossi}, P., \& {Ferrari}, A. 2008,
  \aap, 487, 1

\bibitem[{Boldyrev \& Cattaneo(2004)}]{BC04}
Boldyrev, S., \& Cattaneo, F. 2004, Phys. Rev. Lett., 92, 144501

\bibitem[{{Davis} {et~al.}(2010){Davis}, {Stone}, \& {Pessah}}]{Davis10}
{Davis}, S.~W., {Stone}, J.~M., \& {Pessah}, M.~E. 2010, \apj, 713, 52

\bibitem[{{Fromang}(2010)}]{Fromang10}
{Fromang}, S. 2010, \aap, 514, L5+

\bibitem[{{Fromang} \& {Papaloizou}(2007)}]{Fromang07}
{Fromang}, S., \& {Papaloizou}, J. 2007, \aap, 476, 1113

\bibitem[{{Guan} {et~al.}(2009){Guan}, {Gammie}, {Simon}, \&
  {Johnson}}]{Guan09}
{Guan}, X., {Gammie}, C.~F., {Simon}, J.~B., \& {Johnson}, B.~M. 2009, \apj,
  694, 1010

\bibitem[{{Hawley} {et~al.}(1995){Hawley}, {Gammie}, \& {Balbus}}]{Hawley95}
{Hawley}, J.~F., {Gammie}, C.~F., \& {Balbus}, S.~A. 1995, \apj, 440, 742

\bibitem[{{K{\"a}pyl{\"a}} \& {Korpi}(2010)}]{Kapyla10}
{K{\"a}pyl{\"a}}, P.~J., \& {Korpi}, M.~J. 2010, ArXiv e-prints

\bibitem[{{Lesur} \& {Longaretti}(2007)}]{LL07}
{Lesur}, G., \& {Longaretti}, P. 2007, \mnras, 378, 1471

\bibitem[{{Longaretti} \& {Lesur}(2010)}]{LL10}
{Longaretti}, P., \& {Lesur}, G. 2010, \aap, 516, A51+

\bibitem[{{Mignone} {et~al.}(2007){Mignone}, {Bodo}, {Massaglia}, {Matsakos},
  {Tesileanu}, {Zanni}, \& {Ferrari}}]{PLUTO}
{Mignone}, A., {Bodo}, G., {Massaglia}, S., {Matsakos}, T., {Tesileanu}, O.,
  {Zanni}, C., \& {Ferrari}, A. 2007, \apjs, 170, 228

\bibitem[{{Mignone} \& {Tzeferacos}(2010)}]{Mignone10a}
{Mignone}, A., \& {Tzeferacos}, P. 2010, Journal of Computational Physics, 229,
  2117

\bibitem[{{Mignone} {et~al.}(2010){Mignone}, {Tzeferacos}, \&
  {Bodo}}]{Mignone10b}
{Mignone}, A., {Tzeferacos}, P., \& {Bodo}, G. 2010, Journal of Computational
  Physics, 229, 5896

\bibitem[{{Pessah} {et~al.}(2007){Pessah}, {Chan}, \& {Psaltis}}]{Pessah07}
{Pessah}, M.~E., {Chan}, C., \& {Psaltis}, D. 2007, \apjl, 668, L51

\bibitem[{{Regev} \& {Umurhan}(2008)}]{Regev08}
{Regev}, O., \& {Umurhan}, O.~M. 2008, \aap, 481, 21

\bibitem[{{Schekochihin} {et~al.}(2004){Schekochihin}, {Cowley}, {Taylor},
  {Maron}, \& {McWilliams}}]{Schekochihin04}
{Schekochihin}, A.~A., {Cowley}, S.~C., {Taylor}, S.~F., {Maron}, J.~L., \&
  {McWilliams}, J.~C. 2004, \apj, 612, 276

\bibitem[{{Shakura} \& {Sunyaev}(1973)}]{SS73}
{Shakura}, N.~I., \& {Sunyaev}, R.~A. 1973, \aap, 24, 337

\bibitem[{{Silvers}(2008)}]{Silvers08}
{Silvers}, L.~J. 2008, \mnras, 385, 1036

\bibitem[{{Simon} {et~al.}(2009){Simon}, {Hawley}, \& {Beckwith}}]{Simon09}
{Simon}, J.~B., {Hawley}, J.~F., \& {Beckwith}, K. 2009, \apj, 690, 974

\bibitem[{{Suresh} \& {Huynh}(1997)}]{Suresh97}
{Suresh}, A., \& {Huynh}, H.~T. 1997, Journal of Computational Physics, 136, 83

\bibitem[{Tobias {et~al.}(2011)Tobias, Cattaneo, \& Brummell}]{TCB11}
Tobias, S.~M., Cattaneo, F., \& Brummell, N.~H. 2011, \apj, 728, 153

\bibitem[{Vainshtein \& Kichatinov(1986)}]{VK86}
Vainshtein, S.~I., \& Kichatinov, L.~L. 1986, Journal of Fluid Mechanics (ISSN
  0022-1120), 168, 73

\bibitem[{Vishniac \& Cho(2001)}]{VC01}
Vishniac, E.~T., \& Cho, J. 2001, \apj, 550, 752

\end{thebibliography}
\end{document}